# Deformation of a soft boundary induced and enhanced by enclosed active particles

Wen-de Tian[1,3], Yong-kun Guo[1], Kang Chen[1,3,*] and Yu-qiang Ma[1,2,*]

[1]Center for Soft Condensed Matter Physics & Interdisciplinary Research, College of Physics, Optoelectronics and Energy, Soochow University, Suzhou 215006, China

[2]National Laboratory of Solid State Microstructures and Department of Physics, Nanjing University, Nanjing 210093, China

[3]Kavli Institute for Theoretical Physics China, CAS, Beijing 100190, China

Corresponding authors: tianwende@suda.edu.cn, kangchen@suda.edu.cn; myqiang@nju.edu.cn

## Abstract

We simulate a two dimensional model of a deformable boundary with the confined active particles. The particles tend to accumulate near the boundary and the shape of the boundary deforms upon the enhanced collisions. We find that there are two typical stages of the morphology changing with the increase of self-propelled force on active particles. One is at a small force characterized by the radially inhomogeneous distribution of particles and the suppression of local fluctuations of the circular boundary. The other is at a large force featured by the angularly inhomogeneous distribution of particles and the global shape deformation of soft boundary. The latter two processes are strongly cooperative. We also find different mechanisms in the particle redistribution and opposite force-dependences of the rate of the shape variation at low and high particle area fractions.

**Introduction**

Active motions can lead to rich and intriguing phenomena, e.g. nontrivial fluctuations [1,2,3] and novel structures[4-7] from the microscopic to the macroscopic. Various artificial systems have been designed and fabricated, which allow studying non-equilibrium phenomena under a variety of conditions. [8-13] Active motions play important roles in living systems to maintain large numbers of vital processes such as cargo transport in the crowded environment of cells and the food intake of bacteria[1] . In biology, the microtubules as active matter show a tread-milling behavior due to the polymerization or de-polymerization of their building blocks occurring with the nucleoside triphosphate hydrolysis. This can generate the active pressure on membrane, which is a necessary prerequisite for the shape and motion of cell. [8]

Boundary is ubiquitous in both natural and laboratory active systems. [14-23] The boundary can rectify the motion of active units and cause their accumulation nearby. In return, the reactions from these units can lead the immersed objects with asymmetric boundaries to move directionally [24-27]. Naturally, living units usually operate with soft boundaries such as the cell membrane. It has been shown that the mechanical pressure is not a state variable for the active systems.[28] This can lead to more interesting behaviors for the soft boundary due to the spatially varying pressure. Therefore, understanding the interplay between active units and flexible boundaries is of fundamental importance [29-30].

In this paper, we simply address this problem by studying a simple two dimensional (2D) model, in which self-propelled disks are confined inside a closed deformable boundary. This system is relevant to the cases where active organisms are enclosed by membranes [31]. It can also be realized by adding microscopic self-propelled particles (SPPs) inside a giant vesicle. The particles accumulate near the boundary driven by the active force [14,15] and the collisions by the particles can generate deformations on the boundary. On the one hand, the inhomogeneous aggregation of particles along the boundary leads to its global shape deformation. On the other hand, the trend of particles accumulating at the place of large curvature [20] reinforces the deformation. In the present paper, we are concerned with the variation of the size and shape of the soft boundary and its coupling with the distribution of SPPs. In 2D, the boundary is like a ring polymer, so we use quantities of polymer physics to characterize the behavior of boundary.

**Model and Simulation Methods**

We consider a 2D system where $N_p$ self-propelled particles (treated as disks) are enclosed by a soft boundary. Each particle has a mass, $m$, and diameter, $\sigma$. In addition to the random Brownian motion at temperature $T$, each particle is propelled by a force through its center with the amplitude $F$ with $F=0$ for the passive beads. The orientation of the force, $\hat{\mu}=(\cos\theta,\sin\theta)$, rotates with time due to thermal fluctuation. The soft boundary (without bending energy) is mimicked by a ring composed of $N_b(=1256)$ passive beads connected by harmonic springs. These boundary beads are set to be identical to the active particles except being passive.

The purely repulsive Weeks-Chandler-Andersen (WCA) potential is adopted for the interactions between all non-bonded pairs of particles and beads [32],

$$U_{WCA}(r) = \begin{cases} 4\varepsilon\left[\left(\frac{\sigma}{r}\right)^{12}-\left(\frac{\sigma}{r}\right)^{6}\right]+\varepsilon & r<r_c=\sqrt[6]{2}\sigma \\ 0 & r>r_c \end{cases} \quad (1)$$

$\varepsilon$ is the interaction strength. The harmonic potential between successive beads is $U=k(r-r_0)^2$. Here $k$ is the spring constant; $r$ is the distance between the bonded beads and $r_0$ is the equilibrium bond length. We set $k=50000\,\varepsilon/\sigma^2$ and $r_0=0.25\sigma$ to prevent the active particles from crossing the boundary. Initially, the radius, $R_0$, of the boundary is $50\sigma$ and the enclosed area, $S_0$, is $7850\sigma^2$.

As the traditional polymer systems, Langevin equation is used to describe the motion of active particles and beads of the boundary,

$$m\ddot{\boldsymbol{r}}_i = -\frac{\partial U_i}{\partial \boldsymbol{r}_i} - \gamma \dot{\boldsymbol{r}}_i + F\hat{\boldsymbol{\mu}}_i(t) + \sqrt{2\gamma k_B T}\boldsymbol{\eta}_i(t) \tag{2}$$

$$\dot{\theta}_i = \sqrt{2D_r}\,\xi_i(t) \tag{3}$$

where $\boldsymbol{r}_i$ is the position of the $i$th entity; $\gamma$ is the translational friction coefficient; $D_r$ is the rotational diffusion constant. $\boldsymbol{\eta}_i(t)$ is the Gaussian white noise induced by implicit solvent, which satisfies the fluctuation-dissipation theorem, $\langle \eta_{j,\alpha}(t)\eta_{l,\beta}(t')\rangle = \delta_{jl}\delta_{\alpha\beta}\delta(t-t')$, where $\alpha$ and $\beta$ denote components of Cartesian coordinates. And the rotational noise $\xi_i$ is also Gaussian. Eq. (2) governs the translational motion. For the passive beads, $U_i$ is composed of both non-bonded WCA potentials and bonded harmonic potentials; and only Eq. (2) is required. While for active particles, $U_i$ contains only the non-bonded WCA potentials and the Eq. (3) which depicts the coupled rotational kinetics of the driving direction is necessary.

We use LAMMPS software to perform our simulations [33]. Square box of $200\sigma \times 200\sigma$ with periodic condition in both xy directions is adopted. Reduced units are used in the simulation by setting $m=1$, $\sigma=1$, $k_B T=1$. $\tau = \sqrt{m\sigma^2/k_B T}$ is the corresponding unit time. We set $\varepsilon = 10 k_B T$ and the friction coefficient $\gamma = 10$, which is large enough that the motion of particles and beads is effectively damped. Additionally, we set $D_r = 7.5\times 10^{-4}\,\tau^{-1}$ as an independent parameter similar to the previous work [14,17] for a large persistent length of active particles at a small active force. For every case, it was run by a minimum time of $1\times 10^7 \tau$ with a time step $\Delta t = 0.0025\tau$. The dimensionless active force is in unit of $k_B T/\sigma$.

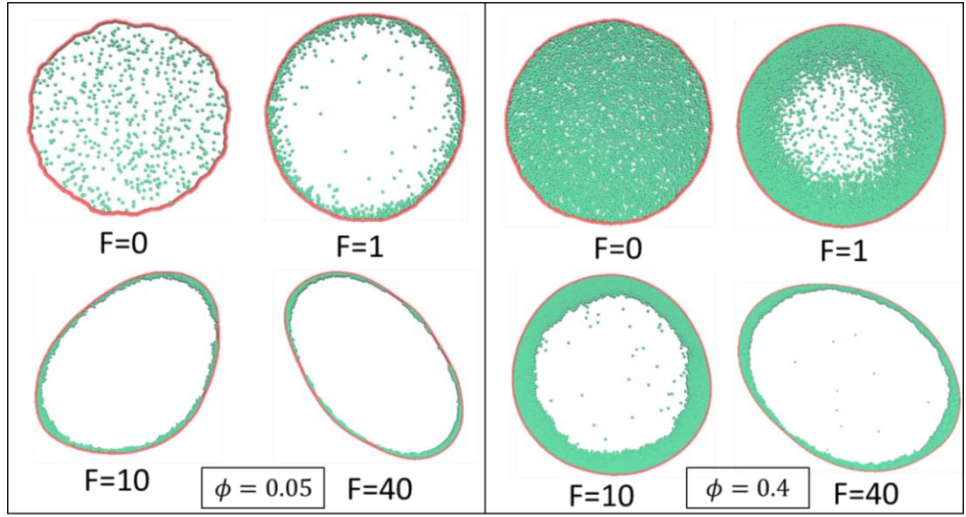

**FIG. 1**. Snapshots of the morphologies for the initial particle area fraction $\phi = 0.05$ (left) and 0.4 (right) at different active forces. The red lines represent the soft bead-spring boundary and the green spheres represent the active particles.

**Results and Discussion**

We consider three initial area fractions of confined SPPs, $\phi = 0.05$ (low), 0.2 (medium), and 0.4 (high), where $\phi \equiv \pi\sigma^2 N_p / 4S_0$. Fig. 1 shows the variation of the typical instantaneous morphologies with the increase of the driving force on the active particles for $\phi = 0.05$ and 0.4. There are two stages in the variation. i) For small active force, the distribution of particles inside the confinement changes from homogeneous to radially inhomogeneous, i.e. particles gradually aggregate around the boundary; the shape of the boundary roughly keeps circular; and apparently the local fluctuation of the boundary is increasingly suppressed at $F$=1.0. ii) For large active force, the distribution of particles becomes both radially and angularly inhomogeneous; the boundary is increasingly elongated and deviating from a circle with the active force; this global shape deformation is the consequence of the inhomogeneous distribution of the particles along the boundary and their trend of accumulating around the places of high curvature [20]. In the following, we quantitatively analyze the variation of the size and shape of the boundary and the inhomogeneous distribution of particles.

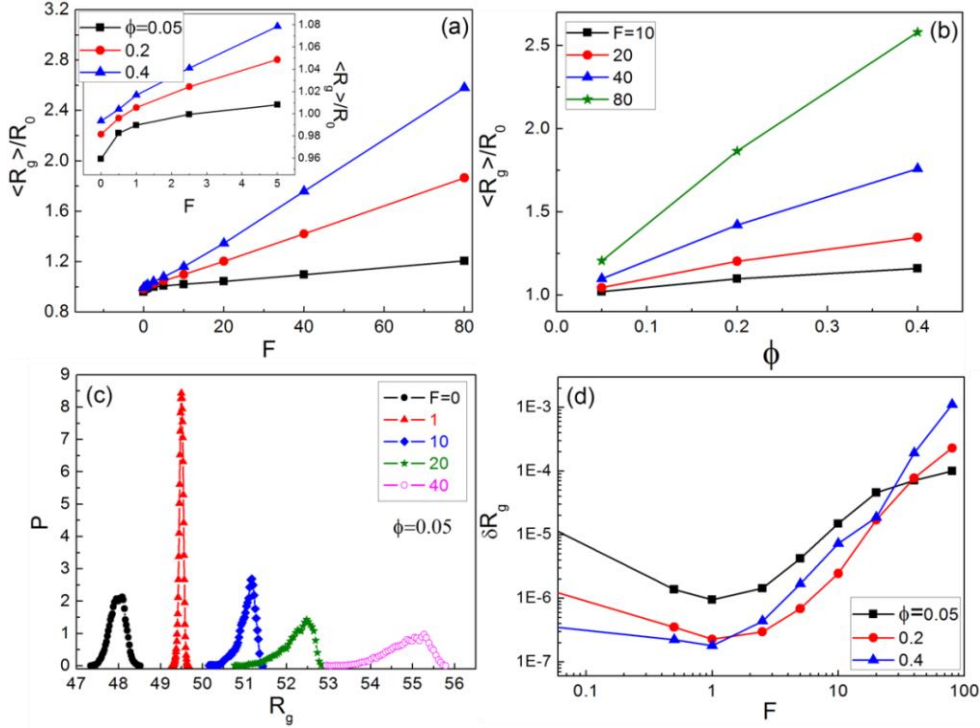

**FIG. 2**. Dimensionless mean radius of gyration of the soft boundary as functions of the active force (a) and initial particle area fraction (b). The inset of (a) zooms in to show the data at small forces. (c) The probability distribution of the radius gyration at various forces for $\phi = 0.05$. (d) The relative fluctuations of the radius of gyration $\delta R_g \equiv \langle \Delta R_g^2 \rangle / \langle R_g \rangle^2$ as a function of the active force.

First, we calculate the mean radius of gyration, $\langle R_g \rangle$, of the boundary, based on the positions of its constituent beads (Fig. 2(a)). At large driving forces ($F \geq 10$), good linear relationships are found between $\langle R_g \rangle$ and $F$ for all $\phi$ s of particles that we have calculated. The slope is larger for higher $\phi$. Fig. 2(b) shows the $\langle R_g \rangle$ as a function of $\phi$ at several (large) forces. The curves only slightly deviate from linear. These close-to-linear behaviors can be explained by a simple crude argument (not considering the inhomogeneous distribution of particles along the boundary) that the expansion force exerting on the boundary is proportional to the strength of the driving force and the number of confined active particles, since thermal forces can be negligible at large driving forces and nearly all the active particles aggregate near the boundary, i.e. $F_{\text{expansion}} \propto \phi F$; the expansion force is balanced by the tension of the boundary, i.e. $F_{\text{expansion}} \propto F_{\text{tension}} \propto k(\Delta R_g)/N_b$; therefore, we have $\Delta R_g \propto \phi F$, i.e. $(R_g - R_0)/R_0$ is a linear function of $F$ or $\phi$. At small $F$ (inset of Fig. 2(a)), we find a crossover from rapid to slow expansion of the boundary with the active force around $F=1$. The initial rapid expansion is attributed to the aggregation of particles to the boundary which effective increase the average number of particles that exert force on the boundary. We calculate the probability distributions of $R_g$ for $\phi=0.05$ and different forces (Fig. 2(c)). The curve of $F=1$ shows the narrowest distribution indicating the strong suppression of the local fluctuation of the boundary due to the increasing tension of boundary. This is further proved by the calculation of the relative fluctuations, $\delta R_g \equiv \langle \Delta R_g^2 \rangle / \langle R_g \rangle^2$ in Fig. 2(d) where minimum fluctuations for all $\phi$s are around $F=1$. The probability distribution of $F \leq 1$ is symmetric but on the contrary the distribution of large forces are asymmetric (Fig. 2(c)). These are the manifestations of the characteristics of the two stages: radial expansion and local fluctuation in stage one and global shape deformation in stage two.

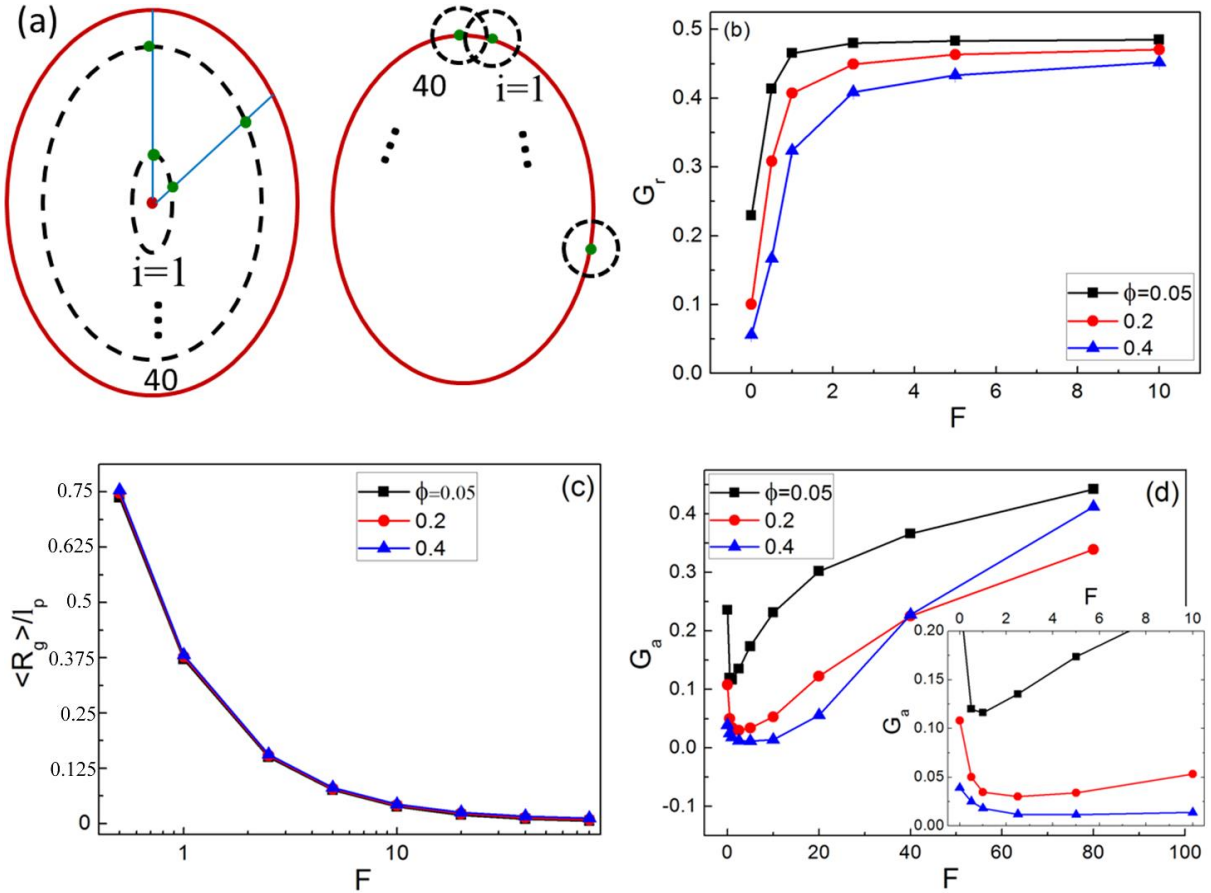

**FIG. 3**. (a) Schematic of partitioning the space radially (left) and angularly along the boundary (right) for calculating the radial and angular Gini coefficients, respectively. (b) Radial Gini coefficient, $G_r$, as a function of the active force. (c) Linear-log plot of the ratio of the mean radius of gyration of the boundary to the persistence length of a free active particle, $\langle R_g \rangle / l_p$, as a function of the active force which crudely quantifies the degree of confinement. (d) Angular Gini coefficient, $G_a$, as a function of the active force; the inset zooms in the data at small forces. (The standard deviations of the data in (b) and (d) are less than 0.09 and the standard errors of the mean less than $4 \times 10^{-4}$.)

To quantitatively describe the radial and angular inhomogeneous distribution of the enclosed particles, we calculated the Gini coefficients [14] $G_r$ and $G_a = \frac{1}{2N^2 \bar{\rho}} \sum_i \sum_j |\rho_i - \rho_j|$ by partitioning the space in radial and angular directions, respectively (see Fig. 3(a) for the schematic). $\bar{\rho}$ is the mean density, $\rho_i$ the number density of particles in the i-th partitioned space, $N$ the total number of partitioned spaces. For the radial Gini coefficient, $G_r$, we chose the center-of-mass of the boundary beads as the center and divided each center-to-bead lines equally into $N = 40$ parts with points and connecting in order the corresponding points on different lines to form inner concentric boundaries which partitioned the space radially. We use the angular Gini coefficient, $G_a$, to capture the inhomogeneous distribution of the particles along the boundary. To this end, we chose $N = 40$ beads on the boundary with an interval of around 30 beads and drew circles with the chosen beads as the centers; the radius is empirically set as $r_c = 0.9L/N$, where $L$ is the perimeter of the boundary. We also tried other choices of $N$ and $L$ which only quantitatively change the results. Fig. 3(b) shows that, for all $\phi$ s, $G_r$ increases significantly when the force $F \leq 1$ and then gradually approaches a plateau. The turning points $F \sim 1$ to 2 mark the end of the stage one in which particles radially redistribute with increasing force. Neglecting the noise in eq. (2), the persistence length of a free active particle $l_p \propto F/\gamma D_r$. The ratio $\langle R_g \rangle / l_p$ can, to some extent, accounts for the degree of confinement [20]. Fig. 3(c) shows a dramatic decrease of the ratio at small force implying a steep transition from weak to strong confinement with the increase of force. The strong confinement causes the formation of the concentration hole in the center (i.e. the bulk particle density approaches zero) [20]. In the regime of stage one, we find the angular Gini coefficient $G_a$ decreases (Fig. 3(d)). We ascribe this novel decrease to the suppression of fluctuations of both the particles and the boundary when particles aggregate near the boundary. At large F, all $G_a$'s increase with force manifesting the enhanced angularly inhomogeneous redistribution of particles (stage two). However, distinct behavior of the crossover from stage one to stage two is found for low and high $\phi$s. For $\phi = 0.05$, $G_a$ increases immediately when $F > 1$, i.e. the crossover from stage one to stage two is sharp without bridging regime. Whereas, $G_a$ keeps small and does not rise appreciably until $F > 5$ for $\phi = 0.2$ and $F > 10$ for $\phi = 0.4$, i.e., there is an evident bridging regime connecting the stage one and stage two for high particle area fractions. In this bridging regime, the whole system expands without significant radial or angular redistribution of the particles.

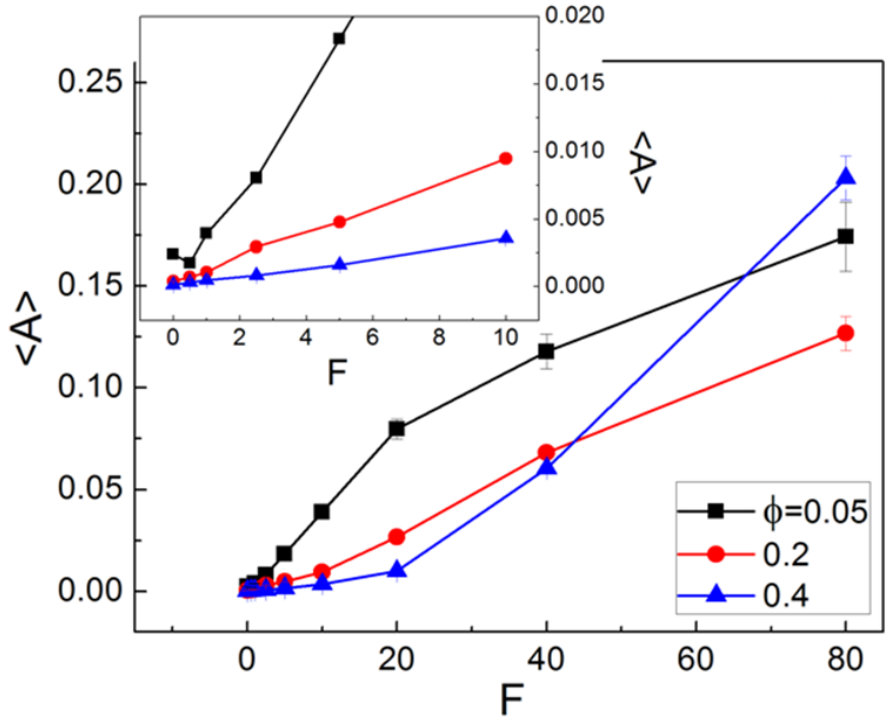

**FIG. 4**. The mean asphericity of the boundary as a function of the active force; the inset zooms in to show the data at small forces.

We calculated the asphericity [34,35] of the boundary to quantify its global shape deformation (deviation from a circle) under various forces and numbers of enclosed particles (Fig .4). The asphericity is defined as:

$$\langle A \rangle = \sum_{i>j} \left\langle \left(\lambda_i - \lambda_j\right)^2 \right\rangle \bigg/ \left\langle \left(\sum_i \lambda_i\right)^2 \right\rangle \qquad (4)$$

where $\lambda_i$ is the i-th eigenvalue of the shape tensor. For our two dimensional boundary, the asphericity is 0 if the shape of the boundary is circular or isotropic. For all $\phi$ s, we find the trends of $\langle A \rangle$ as a function of active force are analogous to those of $G_a$. For example, $\langle A \rangle$ increases dramatically when $F > 1$ in the case of $\phi = 0.05$. We notice that both the curves of $G_a$ and $\langle A \rangle$ for $\phi = 0.05$ show a clear transition from the rapid to slow increase at force around $F = 20$. Examining the trajectories, we find the aggregation layer of particles along the boundary becomes thin and sparse for $\phi = 0.05$ and $F \geq 20$ (Fig. 1), and there is no appreciable adjustment of the thickness of the particle aggregation layers along the boundary upon the further increase of the force, which, otherwise, would contribute to the increase of $G_a$ and $\langle A \rangle$. This slow increase at $F \geq 20$ is then mostly induced by the further expansion of the boundary due to the enhanced pressure by the particles when the active force increases. Clearly it shows a slow to rapid increase for $\phi = 0.2$ and 0.4, where the slow increase corresponds to the bridging regime and the rapid increase corresponds to the stage two. This manifests a strong cooperation between the shape deformation of the boundary and the inhomogeneous angular distribution of particles. With the increase of active forces, more particles arrive at the boundary. The fluctuation of particle collisions with boundary induces the shape fluctuation of boundary, thus the boundary symmetry is broken and high curvature regions appear. It has been demonstrated that this kind of region acts as an attractive center for the active particles [20]. Consequently, more and more particle will accumulate around the high curvature region. Theoretical analysis has shown that pressure on boundary is proportional to the local density of particles, [20] thus further inducing the high curvature. This positive-feedback-like cooperation is enhanced as the active forces increase due to stronger confinement of active particles.

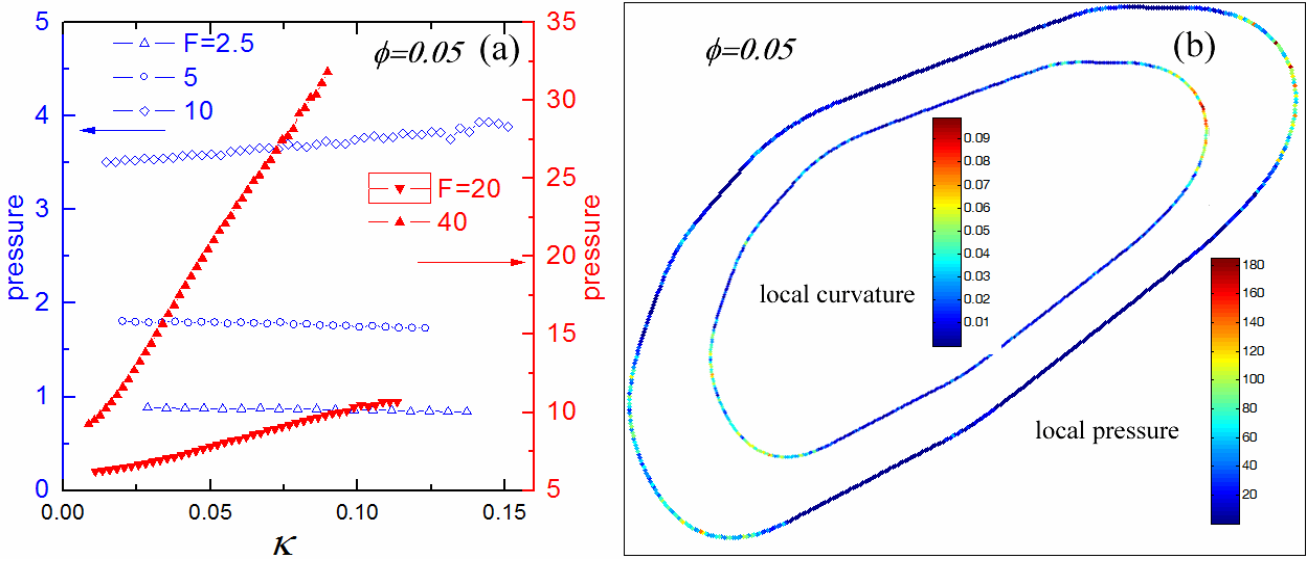

**FIG.5** (a) The average pressure as a function of local curvature, $\kappa$, for $\phi$=0.05. (b) Profile of local curvature (inside) and local pressure (outside) around the boundary for F=80, $\phi$=0.05. The data was extracted from a typical snapshot; the contour of boundary is shown.

To manifest the above mechanism, we calculate the local curvature, $\kappa$, of boundary and the local pressure. Here, $\kappa = \dfrac{|y''|}{(1+y'^2)^{\frac{3}{2}}}$, where (x, y) is the position of boundary point, and $y''$ and $y'$ are the second and first derivatives of x coordinate. The local curvatures are calculated at the beads position along the boundary with discrete form of the derivatives. The local pressure exerted by active particles was evaluated by the local force along the local normal direction, $\hat{n}$, of boundary divided by the local arc length. It can be found that the local pressure decreases slightly with the increase of local curvature at small forces (F≤5.0, Fig.5(a)), corresponding to the radius increase of boundary (Fig.1a). At F≥10, local pressure increases as the local curvature increase (Fig.5(a)). This also can be seen in Fig.5b, which shows a position of high curvature has a high local pressure for F=80, ϕ=0.05. This implies that there exists a strong correlation of local curvature and local pressure at high active forces. The persistent length of particles increases with the increase of active force, then particles enter into the strongly confined region. More and more particles accumulate on the boundary. A fluctuation of local particle density will produce a fluctuation of local pressure, which enhances the local curvature due to the correlation effect. Then the enhanced curvature region attracts more particles.

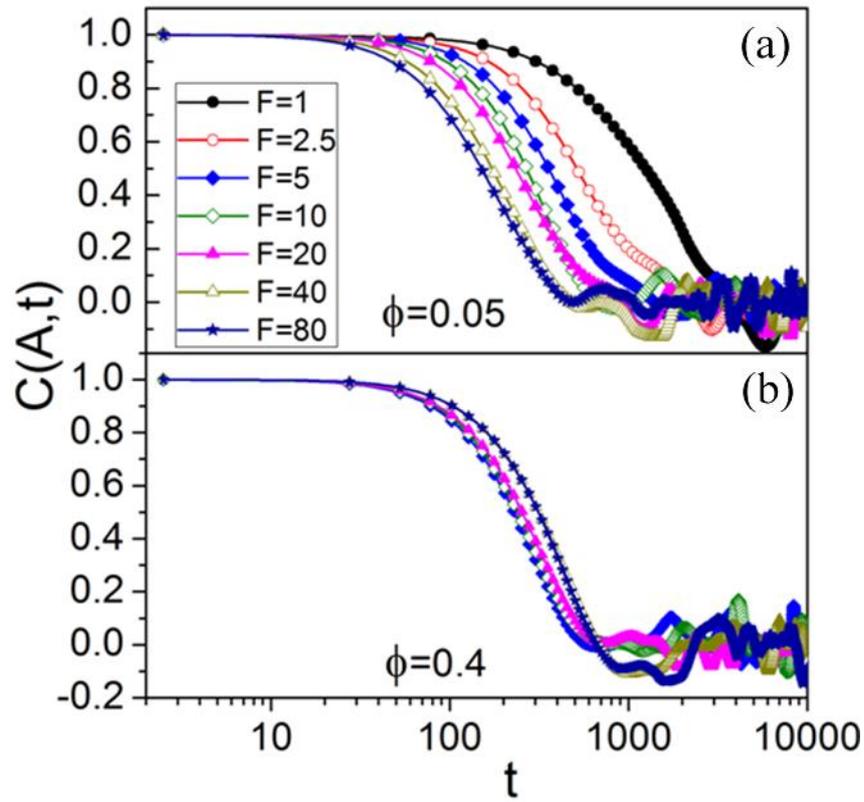

**FIG.6** The decay of the time correlation function of asphericity is shown for $\phi = 0.05$ (b) and 0.4 (c).

Time correlation function of asphericity $C(A,t) \equiv \langle (A(t)-\langle A \rangle)(A(0)-\langle A \rangle) \rangle / (\langle A^2 \rangle - \langle A \rangle^2)$ is a way to quantify the varying rate of the global shape (Fig. 6(a) and (b)). The decay of the correlation function varies with force and initial particle area fraction. We find that the force-dependence of the decay is distinct for low and high $\phi$s. For $\phi = 0.05$, the $C(A,t)$ decays faster (i.e., the shape varies faster) at larger force (Fig. 6(a)). The change of the decaying rate is appreciable. Taking the time at which $C(A,t) = 0.3$ as the characteristic correlation time, we have that the correlation time changes from $t_c = 505$ for $F = 5$ to $t_c = 218$ for $F = 80$. On the contrary, the decay of $C(A,t)$ becomes slower upon the increase of force for $\phi = 0.4$ (Fig. 6(b)), but apparently the change is small. The correlation time changes from $t_c = 312$ for $F = 5$ to $t_c = 418$ for $F = 80$. There are two competing factors in determining the rate of the shape variation: expansion of the boundary and the motility of the particles, both of which are enhanced with the increase of active force. Large size of the boundary slows down the shape variation while high motility of particles leads to the rapid redistribution of them along the boundary and hence the fast shape variation. The expansion of the boundary with active force is weak for $\phi = 0.05$ (Fig. 2(a)) and the enhanced motility of particles dominates and leads to faster decay of $C(A,t)$ at larger force. On the contrary, the expansion of the boundary is significant for $\phi = 0.4$ and causes the slower shape variation at larger force. Notice that the correlation time for $\phi = 0.05$ is larger than that for $\phi = 0.4$ when $F \leq 40$. We examine the cartoons (cf. Supplemental Material movies 1 and 2 [36]) and find that the possible reason is the different mechanisms in the particle redistribution. For $\phi = 0.05$, the layer of particles at the boundary is thin and the particles redistribute mostly through the motion along the boundary. While, the layer of particles at the boundary is thick for $\phi = 0.4$, and besides the motion along the boundary, many particles near the center move across the concentration hole, which can definitely speed up the angular redistribution of the particles.

**Conclusion**

In summary, we study a simplified model of SPPs confined by a soft boundary in two dimensions and focus on the cooperation between shape deformation and particle distribution. We find i) the average size of the boundary varies roughly in a linear relation with the active force and the number of enclosed particles; ii) there are two typical stages in the variation of the morphology with the active force; one is characterized by the radially aggregation of particles to the boundary and suppression of the local fluctuations; the other is characterized by angularly inhomogeneous redistribution of particles and global shape deformation of the boundary; iii) the crossover from stage one to stage two is abrupt for small particle area fractions but a bridging regime of nearly isotropic expansion emerges for medium and high particle area fractions; iv) the global shape deformation of the boundary is strongly coupled with the inhomogeneous angular redistribution of the particles and the fluctuations of the boundary shape can be enhanced by pressure inhomogeneity; v) the global shape varies with time, the rate of which shows opposite dependences on the active force for low and high particle area fractions. This work provides new knowledge and insights to the interplay between a very flexible boundary and the enclosed SPPs.

Clearly, our 2D simulations do not consider the explicit hydrodynamics. Nevertheless we are confident that the main conclusions of our manuscript will remain valid, the dynamics such as diffusion coefficient may be influenced by the effect. Additionally, some realistic systems such as cell cytoskeleton are often immersed in a three-dimensional aqueous solution. The effects of hydrodynamics and dimensions, which need more computational costs and different simulation techniques, could be a direction of our future work.

This work is supported by the National Basic Research Program of China (973 Program) No. 2012CB821500 (K.C. and Y.-q.M.) and the National Natural Science Foundation of China (NSFC) Nos. 21474074 (W.-d.T.), 21374073 (K.C.), and 21574096 (K.C.).